\documentstyle[12pt,A4]{article}
\catcode`\@=11

\@addtoreset{equation}{section}
\def\theequation{\arabic{section}.\arabic{equation}}

\def\@normalsize{\@setsize\normalsize{15pt}\xiipt\@xiipt
\abovedisplayskip 14pt plus3pt minus3pt%
\belowdisplayskip \abovedisplayskip
\abovedisplayshortskip  \z@ plus3pt%
\belowdisplayshortskip  7pt plus3.5pt minus0pt}

\def\small{\@setsize\small{13.6pt}\xipt\@xipt
\abovedisplayskip 13pt plus3pt minus3pt%
\belowdisplayskip \abovedisplayskip
\abovedisplayshortskip  \z@ plus3pt%
\belowdisplayshortskip  7pt plus3.5pt minus0pt
\def\@listi{\parsep 4.5pt plus 2pt minus 1pt
            \itemsep \parsep
            \topsep 9pt plus 3pt minus 3pt}}

\def\underline#1{\relax\ifmmode\@@underline#1\else
        $\@@underline{\hbox{#1}}$\relax\fi}
\@twosidetrue

\relax

\catcode`@=12

\catcode`\@=11

\def\section{\@startsection{section}{1}{\z@}{3.5ex plus 1ex minus
   .2ex}{2.3ex plus .2ex}{\large\bf}}

\def\thesection{\Roman{section}.}

\def\appendix{\setcounter{section}{0}
        \def\thesection{APPENDIX }
        \def\theequation{\Alph{section}.\arabic{equation}}}

\def\FERMIPUB{}

\def\ps@headings{\def\@oddfoot{}\def\@evenfoot{}
\def\@oddhead{\hbox{}\hfill
        \makebox[.5\textwidth]{\raggedright\ignorespaces --\thepage{}--
        \hfill {\rm FERMILAB--Pub--\FERMIPUB}}}
\def\@evenhead{\@oddhead}
\def\subsectionmark##1{\markboth{##1}{}}
}

\catcode`\@=12

\relax

\def\figcap{\section*{Figure Captions\markboth
        {FIGURECAPTIONS}{FIGURECAPTIONS}}\list
        {Fig. \arabic{enumi}:\hfill}{\settowidth\labelwidth{Fig. 999:}
        \leftmargin\labelwidth
        \advance\leftmargin\labelsep\usecounter{enumi}}}
 \relax
\def\tablecap{\section*{Table Captions\markboth
        {TABLECAPTIONS}{TABLECAPTIONS}}\list
        {Table \arabic{enumi}:\hfill}{\settowidth\labelwidth{Table 999:}
        \leftmargin\labelwidth
        \advance\leftmargin\labelsep\usecounter{enumi}}}
 \relax
\def\reflist{\section*{References\markboth
        {REFLIST}{REFLIST}}\list
        {[\arabic{enumi}]\hfill}{\settowidth\labelwidth{[999]}
        \leftmargin\labelwidth
        \advance\leftmargin\labelsep\usecounter{enumi}}}
 \relax

\catcode`\@=11

\def\FERMIPUB{}

\def\ps@headings{\def\@oddfoot{}\def\@evenfoot{}
\def\@oddhead{\hbox{}\hfill
        \makebox[.5\textwidth]{\raggedright\ignorespaces --\thepage{}--
        \hfill {\rm FERMILAB--Pub--\FERMIPUB}}}
\def\@evenhead{\@oddhead}
\def\subsectionmark##1{\markboth{##1}{}}
}

\ps@headings

\relax
\newskip\humongous \humongous=0pt plus 1000pt minus 1000pt
\def\caja{\mathsurround=0pt}
\def\eqalign#1{\,\vcenter{\openup1\jot \caja
        \ialign{\strut \hfil$\displaystyle{##}$&$
        \displaystyle{{}##}$\hfil\crcr#1\crcr}}\,}
\newif\ifdtup

\def\beq{\begin{equation}}
\def\eeq{\end{equation}}

\def\beqn{\begin{eqnarray}}
\def\eeqn{\end{eqnarray}}
\relax

\def\G2{{\; \rm GeV/}c^2}
\def\G{\; \rm GeV}

\def\dotx{\dotx{\dot\overline{x}}}

\relax

\begin{document}
\hbadness=10000
\begin{titlepage}
\nopagebreak
\begin{flushright}

        {\normalsize
 Kanazawa-93-11\\

  September,1993
}\\
\end{flushright}
\vfill
\begin{center}
{\large \bf Minimal String Unification and Hidden Sector \break
in $Z_8$ Orbifold Models }
\vfill
{\bf Tatsuo Kobayashi }

\vspace{1.5cm}
       Department of Physics, Kanazawa University, \\
       Kanazawa, 920-11, Japan \\
\vfill

\end{center}

\vfill
\nopagebreak
\begin{abstract}
We study the minimal supersymmetric standard model derived from the $Z_8$
orbifold models and its hidden sectors.
We use a target-space duality anomaly cancellation so as to investigate
hidden sectors consistent with the MSSM unification.
For the allowed hidden sectors, we estimate the running gauge coupling
constants making use of threshold corrections due to the higher massive modes.
The calculation is important from the viewpoint of gaugino condensations,
which is one of the most promissing mechanism to break the supersymmetry.

\end{abstract}

\vfill
\end{titlepage}
\pagestyle{plain}
\newpage
\voffset = -2.5 cm

Superstring theories are only candidates for unified theories of all the known
interactions.
To contact the \lq measurable' world, we have to derive in a low energy limit
the standard model including recent LEP measurements, which show that all
gauge coupling constants of the model are unified simultaneously at
$M_{\rm GUT}=10^{16}$GeV within the framework of the minimal supersymmetric
standard model (MSSM) [1-4].
Much work has been devoted to obtain the MSSM as string massless spectra.
The string theory implies that all coupling are identical at a string scale
$M_{\rm string}=5.27\times g_{\rm string}\times 10^{17}$GeV
\cite{Kaplunovsky,Derendinger}, where $g_{\rm string}\simeq 1/\sqrt 2$ is the
universal string coupling constant.
This difference between $M_{\rm GUT}$ and $M_{\rm string}$ seems to reject the
possibility of a minimal superstring model which has the same matter fields as
the MSSM.

However this situation could change by threshold corrections due to higher
massive modes.
Threshold corrections have been calculated in the case of the orbifold models
[5-8].
A target-space duality symmetry \cite{Kikkawa,Sakai} plays an important role
in the calculation and becomes anomalous by loop effects.
This anomaly can be cancelled by the Green-Schwarz mechanism \cite{GS} and
threshold effects due to towers of massive modes.
This anomaly cancellation and the unification of the $SU(3)$, $SU(2)$ and
$U(1)_Y$ gauge coupling constants were investigated systematically in
ref.\cite{Ibanez}, which shows that all the $Z_N$ orbifold models except
$Z_6$-II and $Z_8$-I have no candidate for the minimal superstring unification.
Within the framework of the $Z_8$-I orbifold model, explicit search for the
minimal string model was studied in ref.\cite{KKO}.

In addition, effective field theories derived from the superstring theories
have another problem on a mechanism of the SUSY-breaking.
Some realistic SUSY-breaking is expected to occur in a hidden sector.
Ignorance about the hidden sector also makes it difficult to search realistic
models.
A gaugino condensation mechanism is one of the promissing candidates for the
realistic SUSY-breaking with a hiarachy [14-23].
The scale of the condensation $M_{\rm COND}$ and that of the observable
SUSY-breaking $M_{\rm SUSY}$ are related as
$M_{\rm SUSY} \simeq M_{\rm COND}^3/M_{\rm P}^2$, where $M_{\rm P}$ is the
Planck scale.
In order to lead to the SUSY-breaking at 1TeV, the condensation must occur
near by $10^{13}$GeV.
Therefore it is important to study coupling constants of the hidden gauge
groups around the scale, although we have never understood the condensation
mechanism.

In this paper, using the $Z_8$-I orbifold model we investigate all the
possible minimal string models which have the $SU(3)\times SU(2)\times U(1)_Y$
gauge group, three generations, two Higgs particles, their superpartners and
no extra matter except singlets as the observable sector.
Further we study the unification of the gauge coupling constants of $SU(3)$ and
 $SU(2)$ among the models.
We have an ambiguity on a normalization of the $U(1)_Y$ charges.
Hereafter we do not discuss on the $U(1)_Y$ group.
Next we investigate the hidden sectors consistent with the above observable
sector from the viewpoint of the duality anomaly cancellation and analyze
values of their coupling constants at $10^{13.0}$GeV through renormalization
group equations including threshold corrections.

Here we study the $Z_8$-I orbifold models \cite{ZNOrbi,KKKO}.
The 6-dim orbifold is obtained through a division of ${\bf R}^6$ by space
group elements $(\theta,e_b)$, where $e_b$ are vectors spanning an
$SO(9)\times SO(5)$ lattice and $\theta$ is an automorphism of the lattice.
The twist $\theta$ has eigenvalues exp[$2\pi i (1,2,-3)/8$] in a complex basis
($X_i,\tilde X_i$) ($i=1,2,3$).
The orbifold models consist of the string on the 4-dim space-time and the
orbifold, its right-moving superpartner (RNS string) and a left-moving
$E_8\times E'_8$ gauge part, whose momenta $P^I$ ($I=1\sim 16$) span an
$E_8\times E'_8$ lattice.
When we bosonize the RNS part, their momenta span an $SO(10)$ lattice.
The twist $\theta$ is embedded into the $SO(10)$ and $E_8\times E'_8$ lattices
in terms of shifts $v^t$ ($t=1\sim 5$) and $V^I$ ($I=1\sim 16$), respectively.
The shift $v^t$ is obtained as $v^t=(1,2,-3,0,0)/8$ and the shift $V^I$ should
satisfy a condition: $8V^I=0$ (mod $E_8\times E'_8$ lattice).
All possible shifts $V^I$ are shown explicitly in ref.\cite{Gauge}.
Also the vector $e_b$ is embedded into the $E_8\times E'_8$ lattice by a Wilson
 line $a^I_{e_b}$, whose charactaristics depend on the structure of the
orbifold.
Refs.\cite{KO1,KO2} show that the $Z_8$-I orbifold has two indepent Wilison
lines with order two.

Closed strings on the orbifold are classified into untwisted strings and
twisted ones.
Gauge bosons belong to the untwisted sector and their momenta $P^I$ satisfy
$P^IV^I=$integer and $P^Ia^I_{e_b}$=integer.
Massless matter fields of the untwisted sector have the momenta $P^I$
satisfying $8P^IV^I=$1,2,5 (mod 8) and $P^Ia^I_{e_b}$=integer.
The other untwisted states satisfying $P^Ia^I_{e_b}$=integer correspond to
antimatter fields.
We can find $E_N$ ($N=6,7,8$), $SO(2N)$ ($N\leq 8$) and $SU(N)$ ($N\leq 8$) as
 the gauge subgroups through the shift $V^I$ and the Wilson lines
$a^I_e$ in the $Z_8$-I orbifold models \cite{Gauge}.
Further we study in detail combinations of the shift $V^I$ and the Wilson lines
 $a^I_e$ leading to $SU(3) \times SU(2) \times U(1)^5$ as the observable gauge
group.
We assume that some Higgs mechnism breaks undesirable $U(1)$ symmetries
including $U(1)$'s of the hidden sector except $U(1)_Y$.
An explicit analysis on all the possible shifts and Wilson lines shows the
$Z_8$-I orbifold model with the observable gauge group have the number of the
(3,2) untwisted matter fields associated with the $i$-th plane, $N^i_{(3,2)}$
as follows,
$$N^i_{(3,2)}\leq \mbox{(\underline{1,1,0})},
\eqno(1)$$
where the underline represents any permutation of the elements and
$N^i_{(3,2)}\leq (a,b,c)$ implies $N^1_{(3,2)}\leq a$,
$N^2_{(3,2)}\leq b$ and $N^3_{(3,2)}\leq c$ simultaneously.
Similarly we obtain
$$N^i_{(\overline{3},1)}\leq \mbox{(\underline{2,1,1})} \quad {\rm or } \quad
(2,0,2),
\eqno(2)$$
$$N^i_{(1,2)}\leq (\underline{3},0,\underline{1}), \quad
(\underline{2},1,\underline{0}), \quad (1,2,1) \quad {\rm or } \quad (2,0,2),
$$
where $N_{(\overline 3,1)}$ and $N_{(1,2)}$ represent the number of the
$(\overline 3,1)$ and (1,2) untwisted matters.

For the twisted sector, we can find matter fields in $\theta^m$-twisted states
 ($m=1,2,4,5$).
Massless matter fields of the twisted sector satisfy the following condition:
$$h_{KM}+N_{\rm OSC}+c_m-1=0,
\eqno(3)$$
where $h_{KM}$ is a conformal dimension of the $E_8\times E'_8$ gauge part,
$N_{\rm OSC}$ is a number operator and $c_m$ is obtained as
$$ c_m={1\over 2}\sum_{t=1}^3 \left(|mv^t|-{\rm Int}(|mv^t|)\right)
\left(1-|mv^t|+{\rm Int}(|mv^t|)\right),
\eqno(4)$$
where ${\rm Int}(a)$ represents an integer part of $a$.
A representation $\underline{R}$ of the group $G$ contributes to the
conformal dimension as
$$ h_{KM}={C(\underline{R}) \over C(G)+k},
\eqno(5)$$
where $k$ is a level of a Kac-Moody algebra corresponding to $G$ and
$C(\underline{R})$ ($C(G)$) is a quadratic Casimir of the $\underline{R}$
(adjoint) representation, e.g. $C(G)=N$ for $SU(N)$.
Here and hereafter we restrict ourselves to the case where $k$=1.
For example, a (3,2) representation of $SU(3)\times SU(2)$ has the
conformal dimension $h_{KM}=7/12$.
The (3,2) matter fields can be obtained from states oscillated by
$\partial X_i$  with $N_{\rm OSC}=1/8$ in the $\theta$- and $\theta^5$-twisted
sectors, as well as non-oscillated states.
Similarly matter fields with $N$-dim fundamental representations of $SU(N)$
($N=4\sim 8$) can be obtained from states with $N_{\rm OSC}=0,1/8,2/8$ in the
$\theta$- and $\theta^5$-twisted sectors and $N_{\rm OSC}=0,2/8$ in the
$\theta^2$-twisted sector.
However we can not find the above matter fields in the oscillated states of the
 $\theta^4$-twisted sector.
Note that the $\theta$-twisted states with $N_{\rm OSC}=2/8$ are created by two
 oscillators $(\partial X_1)^2$ with $N_{\rm OSC}=1/8$ and an oscillator
$\partial X_2$ with $N_{\rm OSC}=2/8$.
For (3,1) and (1,2) representations of $SU(3) \times SU(2)$, oscillators
corresponding to $N_{\rm OSC}=3/8$ in the $\theta$- and $\theta^5$-twisted
sectors are allowed in addition to the above range of $N_{\rm OSC}$ for the
$N$-dim fundamental representations of $SU(N)$ ($N=4\sim 8$).

In genaral the orbifold models have the duality symmetry.
The effective field theories derived from the string models are invariant
under the following $SL(2,Z)$ transformation of a modulus $T_i$ ($i=1,2,3$)
associated with the $i$-th plane:
$$ T_i \rightarrow {a_iT_i-ib_i \over ic_iT_i+d_i} ,
\eqno(6)$$
with $a_i,b_i,c_i,d_i\in {\bf Z}$ and $a_id_i-b_ic_i=1$.
Under the duality the chiral matter fields $A_\alpha$ transform as follows,
$$ A_\alpha \rightarrow A_\alpha \prod_{i=1}^3(ic_iT_i+d_i)^{n^i_\alpha},
\eqno(7)$$
where $n^i_\alpha$ is called a modular weight \cite{Dixon2,Ibanez,Bailin}.
The untwisted matter fields associated with the $i$-th plane have
$n^j=-\delta^j_i$.
The $\theta$-, $\theta^2$-, $\theta^4$- and $\theta^5$-twisted sectors without
oscillators have $n^i$=$(-7,-6,-3)/8$, $(-6,-4,-6)/8$, $(-4,0,-4)/8$ and
$(-3,-6,-7)/8$, respectively.
An oscillator $\partial X_i$ reduces the corresponding elements of the modular
weight by one and the oscillator $\partial \tilde X_i$ contributes oppositely.
For example the matter fields with the $N$-dim fundamental representation of
$SU(N)$ ($N=4 \sim 8$) can possess the following modular weights:
$$\eqalign{
n^i=& \mbox{(\underline{--1,0,0}), \quad (\underline {--7},--6,\underline {--3}
)/8, \quad
(\underline {--15},--6,\underline {--3})/8, \quad (\underline {--23},--6,
\underline {--3})/8,}\cr
& \mbox{(\underline {--7},--14,\underline {--3})/8, \quad (--6,--4,--6)/8,
\quad
(\underline {--14},--4,\underline {--6})/8, \quad (--4,0,--4)/8.}}
\eqno(8)$$
Similarly we can obtain allowed modular weights for the (3,2),
$(\overline 3,1)$ and (1,2) matter fields of $SU(3)\times SU(2)$, taking into
account the possible values of $N_{\rm OSC}$.

Loop effects make the duality symmetry anomalous.
Its anomaly coefficient for the $i$-th plane is obtained as
$$b'^i_a=-C(G_a)+\sum_{\underline R} T(\underline R)(1+2n^i_{\underline R}),
\eqno(9)$$
where $a$ represents a suffix for a gauge group and $T(\underline R)$ is an
index given as
$T(\underline R)=C(\underline R){\rm dim}(\underline R)/{\rm dim}(G)$, e.g.,
$T(\underline R)=1/2$ for the $N$-dim fundamental representation of $SU(N)$.
This anomaly can be cancelled by two ways.
One is the Green-Schwarz mechanism, which is independent of the gauge groups.
The other is due to the threshold effects.
Since only the $N=2$ supermultiplets contribute to these effects, the
threshold corrections depend on the modulus whose plane is unrotated under some
 $\theta^m$ twist.
Thus for the $Z_8$-I orbifold the corrections depend on $T_2$.
Further for the first and third planes the duality anomaly should be cancelled
only by the Green-Schwarz mechanism.
Therefore we obtain necessary conditions for the anomaly cancellation as
$$ b'^i_{SU(3)}=b'^i_{SU(2)}=b'^i_a, \quad (i=1,3),
\eqno(10)$$
where $b'^i_a$ implies an anomaly coefficient of some hidden sector.
We assign all the possible modular weights to the three (3,2), six
($\overline 3,1$) and five (1,2) matter fields of $SU(3)\times SU(2)$ using
eqs.(1) and (2), and analyze the anomaly cancellation condition for the $SU(3)$
 and $SU(2)$ parts of eq.(10).
We can find 2946 combinations of ($b'^i_{SU(3)},b'^i_{SU(2)}$) satisfying eq.
(10),
up to the values of $b'^2_{SU(3)}$ and $b'^2_{SU(2)}$.

Now we consider running gauge coupling constants including the threshold
corrections.
The one-loop coupling constants $g_a(\mu)$ at a scale $\mu$ are obtained as
$$ {1\over g_a^2(\mu)}={1\over g_{\rm string}^2}+{b_a \over 16\pi^2}
{\rm log}{M_{\rm string}^2 \over \mu^2}-{1\over 16\pi^2}
(b_a^{\prime 2}- \delta_{GS}^2){\rm log}[(T_2+\overline{T}_2)|\eta(T_2)|^4],
\eqno(11)$$
where $\eta(T)$ is the Dedekind function, $\delta_{GS}^2$ is a gauge group
independent GS coefficient and $b_a$ are $N=1$ $\beta$-function coefficients,
$i.e.$, $b_{\rm SU(3)}=-3$ and $b_{\rm SU(2)}=1$.
{}From this renormalization group flow, we can derive a unified scale $M_{a-b}$
of two couplings, $g_a$ and $g_b$ through the following relation:
$$ {\rm log}{M_{a-b} \over M_{\rm string}}={b_{b}^{\prime 2}-
b_{a}^{\prime 2} \over 2(b_{a}-b_{b})}{\rm log}
[(T_2+\overline{T}_2)|\eta(T_2)|^4] .
\eqno(12)$$
Here we discuss the unification of the $SU(3)$ and $SU(2)$ gauge coupling
constants.
They should be unified at $M_{\rm GUT}$ as shown in the measurements.
Note that \break $[(T_2+\overline{T}_2)|\eta(T_2)|^4]$ is always less than one.
To derive $M_{\rm GUT}<M_{\rm string}$ from eq.(12), the anomaly coefficients
must satisfy the following condition:
$$b'^2_{SU(3)}>b'^2_{SU(2)}.
\eqno(13)$$
We obtain the range of $\Delta b'^2 \equiv b'^2_{SU(3)}-b'^2_{SU(2)}$ for
the assignments of the MSSM matter fields allowed by eqs.(10) and (13) as
$1/4\leq \Delta b'^2 \leq 17/2$.
We use $M_{\rm GUT}=10^{16.0}$GeV and $M_{\rm string}=5.27/\sqrt 2 \times
10^{17.0}$GeV so as to get a minimum value $Re T_2=5.5$ in the minimal
string unification derived from the $Z_8$-I orbifold models.

Now we investigate hidden sectors consistent with the observable sector
discussed above, i.e., the possible assignments for the observable MSSM
unification satisfying eqs.(10) and (13).
At first we consider the hidden gauge subgroups which do not have matter
fields with non-trivial representations.
In the case the anomaly coefficients $b'^i_a$ are determined by $C(G_a)$,
i.e., $b'^i_a=-C(G_a)$, where $C(E_8)=30$, $C(E_7)=18$, $C(E_6)=12$ and
$C(SO(2N))=2N-2$.
The Casimir $C(G)$ for all the possible subgroups are listed in the first
column of Table 1.
For these values of $b'^i_a$ we study whether the hidden secotors satisfies
the condition (10) with the observable MSSM unification.
The results are shown in the second column of Table 1.
In this case the groups with $C(G)\geq 14$ are ruled out as the hidden sector
of the MSSM.
Then we obtain the range of $\Delta b'^2$ for the allowed combinations of the
observable and hidden sectors, and the ranges are found in the third column of
 Table 1.
The column also shows the corresponding values of $Re T_2$.
It seems natural that the value of $Re T$ is of order one.
That requires larger values of $\Delta b'^2$.
Hereafter we discuss the case where $\Delta b'^2>3$.
Actually the value $\Delta b'^2=3$ corresponds to $Re T_2\sim 12$.

Next we study the running gauge coupling constants of the hidden sector allowed
 at the previous stage.
Making use of eq.(12) we can easily get a scale where the couplins constants of
 the hidden and the observable $SU(2)$ (or $SU(3)$) gauge groups unify.
Further we use eq.(11) so as to derive the hidden gauge coupling constant at
$10^t$GeV as
$$ \alpha^{-1}_a(t)=\alpha^{-1}_{\rm GUT} -{b_a \over 2 \pi}{\rm log}{10^t
\over M_{\rm GUT}}+{1 \over 2 \pi}\{ b_a-1+{4(b'^2_a-b'^2_{SU(2)}) \over
\Delta b'^2}\}{\rm log} {M_{\rm string} \over M_{\rm GUT}},
\eqno(14)$$
where $\alpha_a=g^2_a/4 \pi$ and $\alpha_{\rm GUT}$ is obtained by the unified
 coupling constant $g_{\rm GUT}$ of $SU(3)$ and $SU(2)$ as
$\alpha_{\rm GUT}=g^2_{\rm GUT}/4 \pi$.
Now we calculate $\alpha_a(t=13.0)$ of the hidden sector which is consistent
with the minimal string unification from the viewpoint of the duality anomaly
cancellation.
We use $\alpha^{-1}_{\rm GUT}=25.7$ at $M_{\rm GUT}=10^{16.0}$GeV \cite{Amaldi}
and $M_{\rm string}=5.27/\sqrt 2 \times 10^{17.0}$GeV to estimate
$\alpha^{-1}_a$ at $10^{13.0}$ GeV.
The results are found in Table 1, where the fourth column shows the allowed
ranges of $\alpha^{-1}_a(t=13.0)$.
If the value of $\alpha_a$ blows up at a higher energy than $10^{13}$GeV, the
fifth column of Table 1 shows the ranges of the scales $M_{\rm blow}$ where
$\alpha^{-1}_a=0$.
The gaugino condensation of SU(3) or SU(4) without non-trivial matter fields
might happen around $10^{13}$GeV, while the larger gauge groups might lead to
the condensation at a higher scale than $10^{13}$GeV.

Now we study the hidden sector of the gauge groups which have the matter fields
 with the non-trivial representations.
Here we restrict ourselves to the $SU(N)'$ gauge groups.
First of all, we consider the hidden sectors which has the gauge subgroup
$SU(2)'$ and one matter field with the doublet representation.
We assign all the possible modular weights to the matter field and calculate
the values of $b'^i_{\rm SU(2)'}$ satisfying eq.(10).
In a way similar to the above discussion we estimate the range of the
$\alpha^{-1}_{\rm SU(2)'}$ at $10^{13.0}$GeV.
We obtain $14.9 \leq \alpha^{-1}_{\rm SU(2)'} \leq 21.7$ for the hidden
$SU(2)'$ coupling constans consistent with the MSSM as the observable sector.
If we consider the $SU(2)'$ gauge group with two doublets, we derive from the
consistency with the MSSM the values of $\alpha^{-1}_{\rm SU(2)'}$ as
$12.5 \leq \alpha^{-1}_{\rm SU(2)'} \leq 32.2$.
When we discuss the $SU(N)'$ ($N>2$) group with matter fields of non-trivial
representations, we have to take into account a gauge anomaly.
The cancellation of the anomaly requires that there appear the same number of
the $N$-dim fundamental representations as their conjugate ones.
For example we study the hidden sector which has the $SU(3)'$ gauge group and
a pair of 3 and $\overline 3$ matter fields.
We obtain $6.7 \leq \alpha^{-1}_{\rm SU(3)'} \leq 26.5$ at $10^{13.0}$GeV for
gauge coupling constant of the hidden sector consistent with the observable
MSSM unification.
Similarly we estimate other types of the hidden sectors consistent with the
observable MSSM unification using eq.(8).
These analyses for the $SU(N)'$ ($N=4,5,6$) gauge subgroups and four or less
pairs of matter fields with $N$-dim fundamental and its conjugate
representations are found in Table 2.
If the gauge coupling constants blow up in some assignmetnts of the modular
weights to the matter fields, the table shows the maximum blow-up scales
$M_{\rm blow}$ (GeV) in parentheses instead of the minimum values of
$\alpha^{-1}$.
We need several matter fields with non-trivial representations unless the
coupling constants of $SU(7)'$ and $SU(8)'$ blow up at higher scale than
$10^{13}$GeV.
The $SU(7)'$ and $SU(8)'$ groups require at least the four and six
pairs of $N$-dim fundamental and its conjugate representations, respectively.
Actually we have a maximum of $\alpha^{-1}_{SU(7)'}(t=13.0)$ equal to 1.4 in
the hidden sector with $SU(7)'$ and the above four pairs, while the model
shows a maximum blow-up scale $M_{\rm blow}$ is equal to $10^{15.2}$GeV.

At last we discuss the case where gauginos of two gauge subgroups condensate.
Refs.\cite{Casas,Casas2} show that two or more condensations lead to more
realistic SUSY-breaking than the unique condensation.
Here we consider the hidden sectors which have the $SU(4)'$ (or $SU(3)'$) gauge
 group without non-trivial representation matter and the $SU(N)'$ gauge group
with one or two pairs of $N$-dim fundamental and its conjugate
representations, and we investigate their consistency with the MSSM
unification.
It is obvious that there never appear simultaneously different gauge groups
without nontrivial matter.
For example we consider the hidden sector which has the $SU(4)'$ gauge group
without nontrivial matter and $SU(4)'$ gauge group with a pair of 4 and
$\overline 4$ matter fields.
Explicit analysis shows that the above combination is allowed only at
$b'^2_{SU(4)'}=-3$ as well as $b'^i_{SU(4)'}=-4$ ($i=1,3$) for the latter
$SU(4)'$.
Further in a way similat to the above we obtain the range of
$\alpha^{-1}_{SU(4)'}$ for the latter $SU(4)'$ as
$6.0<\alpha^{-1}_{SU(4)'}<11.5$ at $10^{13.0}$GeV.
Of course the $SU(4)'$ gauge subgroup without matter gives the same values of
$\alpha^{-1}$ as in Table 1.
The results for the allowed hidden sectors are shown in Table 3, where the
second and third rows show $\alpha^{-1}(t=13.0)$ of the hidden sectors
consistent with $SU(3)'$ and $SU(4)'$ hidden subgroups which have no nontrivial
 matter.
In this case, the given values of $b'^2_{SU(N)'}$ are restricted tightly and
the corresponding values are written in the second and third rows.
The condition due to the duality anomaly cancellation forbids the other
combinations of the $SU(4)'$ (or $SU(3)'$) group without matter and the
$SU(N)'$ ($N\leq 8$) group with the two or less pairs of matter fields which
have the $N$-dim fundamental and its conjugate representations.
For the allowed combinations, the ranges of $\alpha^{-1}$ become narrow
compared with those in Table 2.
Therefore imposition of the two or more gaugino condensations avails to
constrain the hidden sector.

To sum up, we have studied all the possible MSSM's from the $Z_8$-I orbifold
models and their hidden sectors.
The orbifold models rule out the gauge groups with Casimir $C(G)\geq 14$
and no nontrivial matter field as the hidden sector of the observable MSSM.
Further we have estimated the gauge coupling constants of the hidden sectors
using the threshold corrections.
It is interesting to apply the above analyses to the case of the $Z_6$-II
orbifold model, which is another promissing model for the MSSM among the $Z_N$
orbifold models.
Extension to $Z_N \times Z_M$ orbifold models is more complicated,
since three planes of the orbifolds are unrotated under some twist.
However it is important to discuss the orbifold models from the viewpoint of
the above approach.

\newpage
\leftline{\large \bf Note added}
\vspace{0.8 cm}

After almost completion of this work, the author received by S.~Stieberger a
paper \cite{Stieberger}, which show threshold corrections are modified in
 the case where orbifolds are not decomposed into three 2-dim orbifolds,
although they do not discuss the $Z_8$-I orbifold models.
Even if we need a modification on the threshold corrections, the above
discussions
do not change except replacement of the value of $ReT_2$ leading to
$M_{\rm GUT}$.

\vspace{0.8 cm}
\leftline{\large \bf Acknowledgement}
\vspace{0.8 cm}

The author would like to thank N.~Ohtsubo, D.~Suematsu, H.~Kawabe and
K.~Matsubara for useful discussions and also S.~Stieberger for informing him
ref.\cite{Stieberger}.


\newpage
\pagestyle{empty}
\noindent

\begin{center}

{\large Table 1. hidden sectors without matter}\\
\end{center}
The second column shows whether or not each gauge group is allowed as the
hidden sector of the MSSM unification from the viewpont of the duality
anomaly cancellation.
+ and -- imply allowed groups and forbidden ones, respectively.
\begin{center}
\vspace{10mm}
\footnotesize
\begin{tabular}{|c|c|c|c|c|}
\hline
$C(G)$ & Anom. canc. & $\Delta b'^2$ \ ($ReT_2$) & $\alpha^{-1}$ &
$M_{\rm blow}$ \ (GeV)\\ \hline \hline
  2   &     +    & $8 \sim 1/4$ \  ($5.8 \sim 115$) & $13.5 \sim 17.8$ & ---
      \\ \hline
  3   &     +    & $8 \sim 1/4$ \  ($5.8 \sim 115$)& $8.1 \sim 12.0$ & --- \\
      \hline
  4   &     +    & $7 \sim 1/4$ \  ($6.4 \sim 115$)& $2.0 \sim 6.2$  & --- \\
      \hline
  5   &     +    & $7 \sim 1/4$ \  ($6.4 \sim 115$)&  --- &
       $10^{13.0} \sim 10^{13.7}$ \\ \hline
  6   &     +    & $13/2 \sim 1/4$ \ ($6.7 \sim 115$)& --- &
       $10^{13.8} \sim 10^{14.3}$ \\ \hline
  7   &     +    & $13/2 \sim 1/4$ \ ($6.7 \sim 115$)& --- &
       $10^{14.4} \sim 10^{14.9}$ \\ \hline
  8   &     +    & $9/2 \sim 1/4$  \ ($8.9 \sim 115$) &  --- &
        $10^{14.9} \sim 10^{15.2}$ \\ \hline
  10  &     +    & $7/2 \sim 1/4$ \ ($11 \sim 115$)  &  --- & $10^{15.7}$ \\
       \hline
  12  &     +    & $3/2 \sim 1/4$ \ ($21 \sim 115$)  &     --- & --- \\ \hline
  14  &     --   & ---             &     --- & --- \\ \hline
  18  &     --   & ---             &     --- & --- \\ \hline
  30  &     --   & ---             &     --- & --- \\ \hline
\end{tabular}

\vspace{3cm}
{\large Table 2. hidden sectors with matter fields}\\
\end{center}
The second row shows $\alpha^{-1}_{SU(N)'}(t=13.0)$ of the corresponding
hidden gauge subgroups $SU(N)'$ ($N=4,5,6$) with one pair of $N$-dim
fundamental and its conjugate representations.
Similarly the third, fourth and fifth rows show
$\alpha^{-1}_{SU(N)}(t=13.0)$ of the subgroups with two, three and four pairs
of the above representations.
\begin{center}
\vspace{10mm}
\footnotesize
\begin{tabular}{|c|c|c|c|}
\hline
Matter  & $SU(4)$         & $SU(5)$     & $SU(6)$    \\ \hline \hline
1 pair  & $3.3 \sim 11.5$ & $(10^{13.5})\sim 5.8$  & $(10^{14.2})\sim 0.1$ \\
\hline
2 pairs & $3.6 \sim 13.8$ & $(10^{13.3})\sim 8.1$  & $(10^{13.9})\sim 2.4$ \\
\hline
3 pairs & $3.2 \sim 16.2$ & $(10^{13.4})\sim 10.5$ & $(10^{14.2})\sim 4.7$ \\
\hline
4 pairs & $3.5 \sim 18.5$ & $(10^{13.4})\sim 12.8$ & $(10^{14.4})\sim 7.1$ \\
\hline
\end{tabular}

\newpage
{\large Table 3. hidden sectors with two gaugino condensations}\\
\end{center}
In the first row ($N,M$) represents the gauge group $SU(N)'$ with $M$ pairs of
$N$-dim fundamental and its conjugate representations as the hidden sectors.
\begin{center}
\vspace{10mm}
\footnotesize
\begin{tabular}{|c||c|c|c|c|c|}
\hline
($N,M$) & (4,1) & (5,1) & (4,2) & (5,2) & (6,2) \\ \hline
$SU(3)$ & $4.6 \sim 10.1$ & ---  & $7.0 \sim 12.5$ & $0.0\sim 5.4$ & --- \\
 ($b'^2_{SU(N)}$) & ($-5$) &     & ($-4$) & ($-7$) &     \\ \hline
$SU(4)$ & $6.0 \sim 11.5$ & $(10^{13.2})\sim 4.4$ & $5.0 \sim 13.8$ &
$1.3\sim 6.8$ & $(10^{13.8})\sim 0.4$ \\
 ($b'^2_{SU(N)}$) & ($-3$) & ($-6$) & ($-2,-4,-5,-6,-7$) & ($-5$) & $(-7$) \\
\hline
\end{tabular}

\end{center}

\newpage
\begin{thebibliography}{99}


\bibitem{Ellis}
J.~Ellis, S.~Kelley and D.V.~Nanopoulous, Phys.~Lett. {\bf B260} (1991) 131.

\bibitem{Amaldi}
U.~Amaldi, W.~de~Boer and H.~F\"urstenau, Phys.~Lett. {\bf B260} (1991) 447.

\bibitem{Langacker}
P.~Langacker and M.~Luo, Phys.~Rev. {\bf D44} (1991) 817.

\bibitem{Ross}
G.G.~Ross and R.G.~Roberts, Nucl.~Phys. {\bf B377} (1992) 571.

\bibitem{Kaplunovsky}
V.S.~Kaplunovsky, Nucl.~Phys. {\bf B307} (1988) 145.

\bibitem{Derendinger}
J.-P.~Derendinger, S.~Ferrara, C.~Kounnas and F.~Zwirner, Nucl.~Phys.
{\bf B372} (1992) 145.

\bibitem{Dixon}
L.J.~Dixon, V.S.~Kaplunovsky and J.~Louis, Nucl.~Phys. {\bf B355} (1991) 649.

\bibitem{Antoniadis}
I.~Antoniadis, K.S.~Narain and T.R.~Taylor, Phys.~Lett. {\bf B267} (1991) 37.

\bibitem{Kikkawa}
K.~Kikkawa and M.~Yamasaki, Phys.~Lett. {\bf B149} (1984) 357.

\bibitem{Sakai}
N.~Sakai and I.~Senda, Prog.~Theor.~Phys. {\bf 75} (1986) 692.

\bibitem{GS}
M.B.~Green and J.H.~Schwarz, Phys.~Lett. {\bf B149} (1984) 117.

\bibitem{Ibanez}
L.E.~Ib\'a\~nez and D.~L\"ust, Nucl.~Phys. {\bf B382} (1992) 305.

\bibitem{KKO}
H.~Kawabe, T.~Kobayashi and N.~Ohtsubo, preprint Kanazawa-93-08

\bibitem{Nilles}
H.P.~Nilles, Phys.~Lett. {\bf B115} (1982) 193.

\bibitem{Ferrara}
S.~Ferrara, L.~Girardello and H.P.~Nilles, Phys.~Lett. {\bf B125} (1983) 457.

\bibitem{Derendinger2}
J.P.~Derendinger, L.E.~Ib\'a\~nez and H.P.~Nilles, Phys.~Lett. {\bf B155}
(1985) 65.

\bibitem{Dine}
M.~Dine, R.~Rohm N.~Seiberg and E.~Witten, Phys.~Lett. {\bf B156} (1985) 55.

\bibitem{Kounnas}
C.~Kounnas and M.~Porrati, Phys.~Lett. {\bf B191} (1987) 91.

\bibitem{Casas}
J.A.~Casas, Z.~Lalak, C.~Mu\~noz and G.G.~Ross, Nucl.~Phys. {\bf B347} (1990)
243.

\bibitem{Ibanez2}
A.~Font, L.E.~Ib\'a\~nez, D.~L\"ust and F.~Quevedo, Phys.~Lett. {\bf B245}
(1990) 401.

\bibitem{Ferrara2}
S.~Ferrara, N.~Magnoli, T.R.~Taylor and G.~Veneziano, Phys.~Lett. {\bf B245}
(1990) 409.

\bibitem{Ibanez3}
M.~Cveti$\check{\rm c}$, A.~Font, L.E.~Ib\'a\~nez, D.~L\"ust and F.~Quevedo,
Phys.~Lett. {\bf B245} (1990) 401.

\bibitem{Casas2}
B.~de~Carlos, J.A.~Casas and C.~Mu\~noz, preprint CERN-TH.6436/92.

\bibitem{ZNOrbi}
L.~Dixon, J.~Harvey, C.~Vafa and E.~Witten, Nucl.~Phys. {\bf B261} (1985) 678;
Nucl.~Phys. {\bf B274} (1986) 285.

\bibitem{KKKO}
Y.~Katsuki, Y.~Kawamura, T.~Kobayashi, N.~Ohtsubo, Y.~Ono and K.~Tanioka,
Nucl.~Phys. {\bf B341} (1990) 611.

\bibitem{Gauge}
Y.~Katsuki, Y.~Kawamura, T.~Kobayashi, N.~Ohtsubo, and K.~Tanioka,
Prog.~Theor.~Phys. {\bf 82} (1989) 171.

\bibitem{KO1}
T.~Kobayashi and N.~Ohtsubo, Phys.~Lett. {\bf B257} (1991) 56.

\bibitem{KO2}
T.~Kobayashi and N.~Ohtsubo, preprint DPKU-9103 to be published in
Int.J.~Mod.~Phys.~A.

\bibitem{Dixon2}
L.J.~Dixon, V.S.~Kaplunovsky and J.~Louis, Nucl.~Phys. {\bf B329} (1990) 27.

\bibitem{Bailin}
D.~Bailin and A.~Love, Phys.~Lett. {\bf B288} (1992) 263.

\bibitem{Stieberger}
P.~Mayr and S.~Stieberger, preprint MPI-Ph/93-07.

\end{thebibliography}
\end{document}